\documentclass[12pt,a4paper]{article}

\usepackage{amsmath}
\usepackage{amssymb}
\usepackage{amsthm}

\newcommand{\CC}{\mathbb{C}} 

\newcommand{\Hh}{\mathfrak{H}}

\newcommand{\tp}{{}^t}

\theoremstyle{remark}

 \DeclareMathOperator{\im}{Im}

\def\be{\begin{equation}}
\def\ee{\end{equation}}

\begin{document}

\title{A Surprising Relation for the Effective Coupling Constants of $N=2$ Super Yang-Mills Theories}
\author{Marco Matone}\date{}

\maketitle

\begin{center} Dipartimento di Fisica e Astronomia ``G. Galilei'' \\
 Istituto
Nazionale di Fisica Nucleare \\
Universit\`a di Padova, Via Marzolo, 8-35131 Padova,
Italy\end{center}

\maketitle

\begin{abstract} \noindent We show that the effective coupling constants $\tau$ of  supersymmetric gauge theories described by hyperelliptic curves
do not distinguish between the lattices of the two kinds of heterotic string. In particular, the following relation
$$
\Theta_{D_{16}^+}(\tau)=\Theta_{E_8}^2(\tau)
$$
holds. This is reminiscent of the relation, by $T$-duality, of
the two heterotic strings. We suggest that such a relation extends to all curves describing effective supersymmetric gauge theories.

\end{abstract}

\newpage

\noindent
Let us consider the Seiberg and Witten effective coupling constant   \cite{Seiberg:1994rs}\cite{Seiberg:1994aj}
\be
\tau_{ij}={\theta_{ij}\over\pi}+\Big({8\pi i\over g^2}\Big)_{ij} \ .
\label{latauij}\ee
Positivity of $g^2$, which is crucial to guarantee
unitarity of the theory, is a consequence of the Riemann bilinear relations, that
imply
\be
{\rm Im}\,\tau_{ij}>0 \ .
\label{positvita}\ee
We now show that $\tau_{ij}$ satisfies another important relation.
\\

\noindent
The Seiberg-Witten theory, that in the simplest case has been proved in \cite{Bonelli:1996ry},
 is described by hyperelliptic curves \cite{Klemm:1994qs}\cite{Argyres:1994xh}
\be
y^2=\prod_{k=1}^{2g+2}(x-a_i ) \ .
\label{leiper}\ee
The standard example is the case of gauge group $SU(n)$ whose hyperelliptic curves have genus $g=n-1$.
\\

\noindent
A basic result for hyperelliptic curves follows by the polynomial identity
\begin{align}
\Big(\sum_{\{ T\coprod T^c\}} & \prod_{i<j;\, i,j\in T}(a_i-a_j)^2  \prod_{i<j;\, i,j\in T^c}(a_i-a_j)^2\Big)^2 \cr\cr
&= 2^g \sum_{\{ T\coprod T^c\}} \prod_{i<j;\, i,j\in T}(a_i-a_j)^4\prod_{i<j;\, i,j\in T^c}(a_i-a_j)^4 \ ,
\label{larelax}\end{align}
where the sum is over the ${1\over2}\big(^{2g+2}_{\,\, g+1}\big)$ partitions $ T\coprod T^c$ of $\{1,\ldots 2g+2\}$ for which both
$T$ and $T^c$ have $g+1$ elements. Such an identity, that holds for $g\geq1$, and that can be proved
by induction by letting $a_{2g+1}=a_{2g+2}$, has been first derived by Poor \cite{Poor}. Using the Thomae formula and some theta identities, this leads
to the following identity for all hyperelliptic Riemann surfaces \cite{Poor}
\be
F_g(\tau)=0 \ ,
\label{lidentitas}\ee
where $F_g(Z)$ is the modular form of weight 8
\be F_g(Z)=2^g \sum_{\delta\hbox{
even}}\theta^{16} [\delta](0,Z)- \bigl(\sum_{\delta\hbox{
even}}\theta^{8}[\delta](0,Z)\bigr)^2 \ ,
\label{nerbusaasolata}\ee
and $Z$ is an arbitrary element of the Siegel upper half-space
$$\Hh_g:=\{Z\in M_g(\CC)\mid \tp Z=Z,\im Z>0\} \ .$$

\noindent
It turns out that $F_g$ is proportional to the difference between the theta series associated to the even
unimodular lattices $E_8\oplus E_8$ and $D_{16}^+$
\be\label{difference}
F_g(Z)=2^{2g}(\Theta_{D_{16}^+}(Z)-\Theta_{E_8}^2(Z))\ . \ee
Let $\mathcal{I}_g$ be the closure of the locus of Riemann period
matrices in $\Hh_g$. $F_4$
is the Schottky-Igusa form \cite{Igusauno,IgusaSc} and the
irreducible variety in $\Hh_4$ defined by $F_4=0$ is $\mathcal{I}_4\subset\Hh_4$. This explicitly solves the Schottky problem for $g=4$.
\\

\noindent
A consequence of the above result is that the effective coupling constants of $N=2$ super Yang-Mills theories with hyperelliptic curves satisfy the relation \\
\be
\Theta_{D_{16}^+}(\tau)=\Theta_{E_8}^2(\tau) \ .
\label{chebella}\ee
\\
This means that the effective coupling constants of effective supersymmetric gauge theories do not distinguish between the two lattices defining
the two kinds of heterotic string, the heterotic $SO(32)$ and the heterotic $E_8\times E_8$ \cite{Gross:1984dd}.
This suggests a possible new connection between string and SYM theories. 
\\

\noindent
It would be then interesting to understand if (\ref{chebella})
extends to all curves describing effective supersymmetric gauge theories. An interesting example is provided by the Gaiotto curves \cite{Gaiotto:2009we}
\be
y^2={{\cal P}_{2n+2}(x)\over x^2(x-x_1)^2\cdots (x-x_n)^2(x-1)^2} \ ,
\label{Gaiotto}\ee
where the $x_i$'s denote the punctures, ${\cal P}_{2n+2}(x)$ is a polynomial of degree $(2n+2)$ depending on the Coulomb branch $u_i$, on the masses and on the gauge couplings $q_i=x_i/x_{i+1}$. Recently, in \cite{Ashok:2015gfa}, using the following analogous of the relation in \cite{Matone:1995rx}
\be
U_i=q_i{\partial F\over \partial q_i} \ ,
\label{like}\ee
the functional dependence of the gauge invariant modulus $U_i=\langle {\rm Tr}\, \Phi_i^2\rangle$ on the $a_i$'s has been obtained. Integrating (\ref{like}) with respect to $\log q_i$ one obtains
the prepotential $F$ \cite{Ashok:2015gfa}. In this way one may compare such an expression with the Nekrasov's prepotential \cite{Nekrasov:2002qd}. In turn, all this is related to the recent work
\cite{Marshakov:2013lga}\cite{Gavrylenko:2013dba},
where it has been proposed to identify the $U_i$'s
with the residues of the quadratic differential $y^2(x)$
at the punctures.
\\

\noindent
We now provide some evidence that (\ref{chebella}) holds for supersymmetric gauge theories described by a prepotential. We first insist on the crucial role of duality. Let us consider the case of gauge group
$SU(2)$.
It is immediately verified that the theory has a basic invariance under the translation of the effective coupling constant. This means that the gauge invariant modulus $u$ has the invariance
$$
u(\tau+2)=u(\tau) \ .
$$
Denote by $T$ the translation operator by a unit. On the other hand, the theory admits a dual representation. This means that the same theory can be expressed in terms of dual variables, in particular in terms of
$$
\tau_D=-{1\over\tau} \ .
$$
Denote by $S$ the inversion operator.
Since the dual theory has the same formal properties of the original formulation, it follows that it also remains invariant under $T^2$, that is
$$
u(\tau_D+2)=u(\tau_D) \ .
$$
It follows that the trivial $T^2$ symmetry, together with the existence of a dual formulation, leads
to the fundamental symmetry group generated by
\begin{align}
u(T^2\tau)&=u(\tau) \ ,\cr\cr
u(S^{-1}T^2S\tau) &= u(S^{-1} T^2\tau_D)= u(S^{-1}\tau_D)=u(\tau) \ .
\end{align}
Since acting iteratively with $T^2$ and $S^{-1}T^2S$ we get $\Gamma(2)$, it follows that this is the symmetry group of $N=2$ SYM with gauge group $SU(2)$. This solves the theory and
leads to
the relation \cite{Matone:1995rx}
\be
u=\pi i \Big(F-{a\over2}{\partial F\over \partial a}\Big) \ .
\label{lr}\ee
A similar construction should lead to a short proof for more general cases. What is important here, is just the existence of a dual formulation of the theory and translation invariance.
Existence of the prepotential, the above relation and the existence of a Riemann period matrix, are strictly related facts. This identifies a particular class of Riemann surfaces, namely the ones with period matrix
$$
\tau_{ij}={\partial^2 F\over\partial {a_i}\partial {a_j}}\ .
$$
\\

\noindent  We note that the presence of punctures in Gaiotto curves may be considered as a suitable limit procedure corresponding to the
degeneration of the moduli space of hyperelliptic curves. However, as a direct check one should apply the polynomial identity (\ref{larelax}) to the polynomials of Gaiotto's curves, identifies the period matrix $\tau_{ij}$ by (\ref{like}), and then using the Thomae formulas to verify whether (\ref{chebella}) holds in this case. More generally, one should investigate, using the inversion method in \cite{Ashok:2015gfa},
if all effective supersymmetric theories admitting the relation
(\ref{lr}) may correspond to effective coupling constants satisfying (\ref{chebella}).
\\

\noindent
We conclude observing that a related approach to check (\ref{chebella}) is to use the M-theory construction \cite{Witten:1997sc}\cite{Isidro:1998vh}. We also note that since the underlying theory of Riemann surfaces is the uniformization theory, it follows that Liouville theory should emerge naturally in such an investigation. The role of Liouville theory in supersymmetric Yang-Mills theories has been observed in
\cite{Matone:1995rx} and \cite{Bonelli:1996ry}. Subsequently, it has been shown that Seiberg-Witten theory can be formulated on the moduli space of punctured spheres, whose geometry is described by the Liouville action \cite{Bertoldi:2004cc}. Interestingly, related structures also appear in the AGT correspondence
\cite{Alday:2009aq}\cite{Alday:2009fs}.

\section*{Acknowledgements} It is a pleasure to thank  Samuel Grushevsky, Daniel Ricci-Pacifici, Giulio Pasini, Paolo Pasti, Riccardo Salvati-Manni,
 Dmitri Sorokin and Roberto Volpato for interesting
discussions.

\newpage

\bibliographystyle{amsplain}

\begin{thebibliography}{99}

\bibitem{Seiberg:1994rs}
  N.~Seiberg and E.~Witten,
  Nucl.\ Phys.\ B {\bf 426} (1994) 19.

\bibitem{Seiberg:1994aj}
  N.~Seiberg and E.~Witten,
  Nucl.\ Phys.\ B {\bf 431} (1994) 484.

\bibitem{Bonelli:1996ry}
  G.~Bonelli, M.~Matone and M.~Tonin,
  Phys.\ Rev.\ D {\bf 55} (1997) 6466.

\bibitem{Klemm:1994qs}
  A.~Klemm, W.~Lerche, S.~Yankielowicz and S.~Theisen,
  Phys.\ Lett.\ B {\bf 344} (1995) 169.

\bibitem{Argyres:1994xh}
  P.~C.~Argyres and A.~E.~Faraggi,
  Phys.\ Rev.\ Lett.\  {\bf 74} (1995) 3931.


\bibitem{Poor} C.~Poor, Schottky's form for and the hyperelliptic locus, {\it Proc. Amer. Math. Soc.} {\bf 124} (1996), 1987-1991.


\bibitem{Igusauno}
J.-I.~Igusa, Schottky's invariant and quadratic forms, {\it E. B.
Christoffel} (Aachen/Monschau, 1979), 352-362, Birkh\"auser, Mass., 1981.

\bibitem{IgusaSc}J.-I.~Igusa, On the irreducibility of Schottky's
divisor, {\it J. Fac. Sci. Univ. Tokyo Sect. IA Math.} {\bf 28}
(1981), 531-545.


\bibitem{Gross:1984dd}
  D.~J.~Gross, J.~A.~Harvey, E.~J.~Martinec and R.~Rohm,
  Phys.\ Rev.\ Lett.\  {\bf 54} (1985) 502;
  Nucl.\ Phys.\ B {\bf 256} (1985) 253;
  Nucl.\ Phys.\ B {\bf 267} (1986) 75.


\bibitem{Gaiotto:2009we}
  D.~Gaiotto,
  JHEP {\bf 1208} (2012) 034.

\bibitem{Ashok:2015gfa}
  S.~K.~Ashok, M.~Bill\'o, E.~Dell'Aquila, M.~Frau, R.~R.~John and A.~Lerda,
  arXiv:1502.05581 [hep-th].

\bibitem{Matone:1995rx}
  M.~Matone,
  Phys.\ Lett.\ B {\bf 357} (1995) 342;
  Phys.\ Rev.\ D {\bf 53} (1996) 7354.

\bibitem{Nekrasov:2002qd}
  N.~A.~Nekrasov,
  Adv.\ Theor.\ Math.\ Phys.\  {\bf 7} (2004) 831.

\bibitem{Marshakov:2013lga}
  A.~Marshakov,
  JHEP {\bf 1307}, 068 (2013).

\bibitem{Gavrylenko:2013dba}
  P.~Gavrylenko and A.~Marshakov,
  JHEP {\bf 1405} (2014) 097.


\bibitem{Witten:1997sc}
  E.~Witten,
  Nucl.\ Phys.\ B {\bf 500} (1997) 3.

\bibitem{Isidro:1998vh}
  J.~M.~Isidro,
  Nucl.\ Phys.\ B {\bf 539} (1999) 379.
 

\bibitem{Bertoldi:2004cc}
  G.~Bertoldi, S.~Bolognesi, M.~Matone, L.~Mazzucato and Y.~Nakayama,
  ``The Liouville geometry of N = 2 instantons and the moduli of punctured spheres,''
  JHEP {\bf 0405} (2004) 075,
 [hep-th/0405117].
  
  \bibitem{Alday:2009aq}
  L.~F.~Alday, D.~Gaiotto and Y.~Tachikawa,
  Lett.\ Math.\ Phys.\  {\bf 91} (2010) 167.

\bibitem{Alday:2009fs}
  L.~F.~Alday, D.~Gaiotto, S.~Gukov, Y.~Tachikawa and H.~Verlinde,
  JHEP {\bf 1001} (2010) 113.
  
 
\end{thebibliography}

\end{document}